\documentstyle[amssymb]{amsart}

\def\beq{\begin{equation}}
\def\eeq{\end{equation}}

\def\bq{\begin{quote}}
\def\eq{\end{quote}}

\def\lqq{\lq \lq }
\relax

\theoremstyle{definition}

\newcommand{\be}{\begin{equation}}
\newcommand{\ee}{\end{equation}}
\begin{document}

\title[Operator identities \dots]
{Operator identities, representations of algebras and the problem of normal
ordering}
\author{ Alexander Turbiner and Gerhard Post}
\address
{Institute for Theoretical and Experimental Physics,
Moscow 117259, Russia (A.T., leave on absense)\quad {\it and}
Department of Applied Mathematics, University of Twente, P.O. Box 217,
\quad \quad 7500 AE  Enschede, The Netherlands (G.P.)}
\email{ TURBINER@@CERNVM or TURBINER@@VXCERN.CERN.CH \quad \quad
{\it and}
 POST@@MATH.UTWENTE.NL}
\curraddr{Math. Dept., Case Western Reserve Univ., Cleveland, Ohio 44106
and Physique Theorique, CRN and University Louis Pasteur, Strasbourg, F-67037,
France}
\thanks{Supported in Part by a CAST grant of the US National Academy of
Sciences}

\date{}
\maketitle
\begin{abstract}
Families of operator identities related to certain powers of
positive root generators of (super) Lie algebras of first-order
differential operators and $q$-deformed algebras of first-order
finite-difference operators are presented. It is shown that those identities
once rewritten in terms of creation/annihilation operators lead
to a simplification of the problem of the normal ordering in the second
quantization method.
\end{abstract}

The method of the second quantization is one of the main tools
in quantum field theory and statistical mechanics. One of the tedious
problems appeared in the appications of this method is
the problem of normal ordering. This Note is devoted to a description
of certain infinite families of relations between creation/annihilation
operators, which can simplify the problem of the normal ordering.
Those relations occur as a consequence of the existence of
finite-dimensional representations of semi-simple Lie algebras.

1. The following {\it operator identity} holds
\label{e1}
\begin{equation}
{(J^+_n)}^{n+1} \equiv (x^2 \partial_x - n x)^{n+1} =
x^{2n+2}\partial_x^{n+1} , \partial_x \equiv {d \over dx}, n=0,1,2,\ldots
\end{equation}
The proof is straightforward:
\begin{enumerate}
\item[(i)] the operator ${(J^+_n)}^{n+1}$
annihilates the space of all polynomials of degree not higher than $n$,
${\cal P}_n(x)=Span\{x^i: 0 \leq i \leq n\}$;
\item[(ii)] in general, an $(n+1)-$th order linear differential
operator annihilating ${\cal P}_n(x)$ must have the form
$B(x)\partial_x^{n+1}$, where $B(x)$ is an arbitrary function and
\item[(iii)]
since ${(J^+_n)}^{n+1}$ is a graded operator, deg$(J^+_n)=+1$,
\footnote{so $J^+_n$ maps $x^k$ to a multiple of $x^{k+1}$ }
deg$({J^+_n})^{n+1}=n+1$, hence $B(x)=b x^{2n+2}$ while clearly the
constant $b=1$.
\end{enumerate}
It is worth noting that taking the degree in (1) different from
$(n+1)$, the l.h.s. in (1) will contain immideately all derivative terms
from zero up to $(n+1)$-th order.

The identity has a Lie-algebraic interpretation.
The operator $(J^+_n)$ is the positive-root generator of the algebra $sl_2$
of first-order differential operators (the other $sl_2$-generators are
$J^0_n = x \partial_x - n/2 \ , J^-_n = \partial_x $). Correspondingly, the
space ${\cal P}_n(x)$ is nothing but the $(n+1)$-dimensional
irreducible representation of $sl_2$. The identity (1) is a
consequence of the fact that ${(J^+_n)}^{n+1}=0$ in matrix representation.

Another Lie-algebraic interpretation of (1) is connected with occurence of
some relations between the elements of the universal enveloping algebra
of the one-dimensional Heisenberg algebra $\{ P, Q, 1\}$.
Once $[P,Q]=1$, then:
\label{e2}
\begin{equation}
(Q^2 P - n Q)^{n+1} = Q^{2n+2} P^{n+1}, \ n=0,1,2,\ldots
\end{equation}
Now let us introduce generators $a=2^{1/2} P$ and $a^+ = 2^{1/2} Q$.
Then (2) takes the form
\label{e3}
\begin{equation}
(a^+a^+ a - 2n a^+)^{n+1} = {(a^+)}^{2n+2} a^{n+1}, \ n=0,1,2,\ldots
\end{equation}
Definitely, one can interpret the operators $a^+,a$ as
creation/annihilation operators, respectively.
One can name (3) {\it the first ordering formula}. Of course, those operators
can be realized in the standard way: $a^+=\partial_x+x$ and $a=\partial_x-x$.

There exist other algebras of differential or finite-difference
operators (in more than one variable), which admit a
finite-dimensional representation.  This leads to more general
and remarkable operator identities and hence to ordering formulas.

2. The Lie-algebraic interpretation presented above allows us to generalize
(1) for the case of differential operators of several variables,
taking appropriate degrees of the highest-positive-root generators of
(super) Lie algebras of first-order differential operators, possessing
a finite-dimensional invariant sub-space (see e.g.\cite{t1}).
First we consider
the case of $sl_3$. There exists a representation of $sl_3(\bold C)$ as
differential operators on $\bold C^2$. One of the generators is
\[ J^1_2 (n)= x^2 \partial_x\ +\ xy \partial_y - n x \]
The space ${\cal P}_n(x,y)=Span\{x^iy^j: 0 \leq i+j \leq n\}$ is a
finite-dimensional representation for $sl_3$, and due to the fact $(J^1_2
(n))^{n+1}=0$ on the space ${\cal P}_n(x,y)$, hence we arrive at
\label{e4}
\begin{eqnarray}
{(J^1_2 (n))}^{n+1} = (x^2 \partial_x\ +\ xy \partial_y - n x)^{n+1} =
\nonumber \\
\sum_{k=0}^{k=n+1} {n+1 \choose k} x^{2n+2-k}y^k \partial_x^{n+1-k}
\partial_y^k  \ ,
\end{eqnarray}
This identity is valid for $y \in \bold C$ (as described above), but also if
$y$ is a Grassmann variable, i.e. $y^2=0$ \footnote{In this case just
two terms in the l.h.s. of (4) survive.}. In the last case, $J^1_2 (n)$ is
a generator of $osp(2,2)$, see \cite{t1}.

In general taking $sl_k$ instead of $sl_3$, the following
operator identity holds
\label{e5}
\begin{eqnarray}
{(J^{k-2}_{k-1} (n))}^{n+1} \equiv (x_1\sum_{m=1}^{k} (x_m \partial_{x_m}\
- n))^{n+1} =
\nonumber \\
x_1^{n+1}\sum_{j_1+j_2+\ldots+j_k=n+1} C^{n+1}_{j_1,j_2,\ldots,j_k}
x_1^{j_1}x_2^{j_2}\ldots x_k^{j_k} \partial_{x_1}^{j_1}
\partial_{x_2}^{j_2}\ldots \partial_{x_k}^{j_k}   \ ,
\end{eqnarray}
where $C^{n+1}_{j_1,j_2,\ldots,j_k}$ are the generalized binomial
(multinomial) coefficients.
If $x \in \bold C^k$, then $J^{k-2}_{k-1} (n)$ is a generator of the algebra
$sl_{k+1}(\bold C)$ \cite{t1}, while some of the variables $x'$s are Grassmann
ones, the operator $J^{k-2}_{k-1} (n)$ is a generator of a certain super
Lie algebra of first-order differential operators. The operator in l.h.s.
of (5) annihilates the linear space of polynomials
${\cal P}_n(x_1,x_2,\ldots x_k)=Span\{x_1^{j_1}x_2^{j_2}\ldots x_k^{j_k} :
 0 \leq j_1+j_2+\ldots+j_k \leq n\}$.

Denoting $Q_m=x_m$ and $P_m=\partial_{x_m}$, one can make the following
statement. Once the operators $Q_m,P_m$ are the generators of $k$-dimensional
Heisenberg algebra:
\[
[P_m,Q_l]=\delta_{ml}
\]
then
\label{e6}
\begin{eqnarray}
(Q_1\sum_{m=1}^{k} (Q_m P_m - n))^{n+1} =
\nonumber \\
Q_1^{n+1}\sum_{j_1+j_2+\ldots+j_k=n+1} C^{n+1}_{j_1,j_2,\ldots,j_k}
Q_1^{j_1}Q_2^{j_2}\ldots Q_k^{j_k}
P_1^{j_1}P_2^{j_2}\ldots P_k^{j_k}   \ ,
\end{eqnarray}
(cf.(2)).
Introducing new operators $a_m=2^{1/2} P_m$ and
$a_m^+=2^{1/2} Q_m$, we arrive at
\label{e7}
\begin{eqnarray}
(a_1^+\sum_{m=1}^{k} (a_m^+ a_m - 2 n))^{n+1} =
\nonumber \\
{(a_1^+)}^{n+1}\sum_{j_1+j_2+\ldots+j_k=n+1} C^{n+1}_{j_1,j_2,\ldots,j_k}
{(a_1^+)}^{j_1}{(a_2^+)}^{j_2}\ldots {(a_k^+)}^{j_k} a_1^{j_1}
a_2^{j_2}\ldots a_k^{j_k}   \ ,
\end{eqnarray}
(cf.(3)). As well as before one can consider a standard representation of
the operators $a_k^+=\partial_{x_k}+x_k,\ a_k=\partial_{x_k}-x_k$ as
 creation/annihilation operators, respectively. One can name (7) {\it the
 $k$-th ordering formula.}

3. The above-described family of operator identities (1) can be generalized
for the case of finite-difference operators with the Jackson symbol, $D_x$
(see e.g. \cite{e})
\[ D_x f(x) = {{f(x) - f(q^2x)} \over {(1 - q^2) x}} + f(q^2x) D_x\]
instead of the ordinary derivative. Here, $q$ is an arbitrary complex
number. The following operator identity holds
\label{e8}
\begin{equation}
{(\tilde J^+_n)}^{n+1} \equiv  (x^2 D_x - \{ n \}  x)^{n+1} =
q^{2n(n+1)}x^{2n+2} D^{n+1}_x , n=0,1,2,\ldots
\end{equation}
(cf.(1)), where $\{n\} = {{1 - q^{2n}}\over {1 - q^2}}$ is so-called
$q$-number.  The operator in the r.h.s. annihilates the space
${\cal P}_n(x)$. The proof is similar to the proof of the identity (1).

{}From algebraic point of view the operator $\tilde J^+_n$ is the generator
of a $q$-deformed algebra $sl_2(\bold C)_q$ of first-order
finite-difference operators on the line: \linebreak
$\tilde J^0_n = \ x D - \hat{n},\ \tilde J^-_n = \ D $, where
$\hat n \equiv {\{n\}\{n+1\}\over \{2n+2\}}$ (see \cite{ot} and
also \cite{t1}), obeying the commutation relations
\label{e9}
\[
q^2 \tilde  j^0\tilde  j^- \ - \ \tilde  j^-\tilde  j^0 \
= \ - \tilde  j^-
\]
\begin{equation}
 q^4 \tilde  j^+\tilde  j^- \ - \ \tilde  j^-\tilde  j^+ \
= \ - (q^2+1) \tilde  j^0
\end{equation}
\[
\tilde  j^0\tilde  j^+ \ - \ q^2\tilde  j^+\tilde  j^0 \ = \  \tilde  j^+
 \]
($\tilde j$'s are related with $\tilde J$'s through some multiplicative
factors). The algebra (9) has the linear space ${\cal P}_n(x)$ as a
finite-dimensional representation.

Evidently, the identity (8) has more general meaning like the identity (1).
Once two operators $\tilde P ,\tilde Q$ obey a condition
$\tilde P \tilde Q - q^2 \tilde Q \tilde P=1$, then
\label{e10}
\begin{equation}
  ({\tilde Q}^2\tilde P - \{ n \} \tilde Q)^{n+1} =
q^{2n(n+1)}{\tilde Q}^{2n+2} {\tilde P}^{n+1} , n=0,1,2,\ldots
\end{equation}
(cf.(2)).

An attempt to generalize (4) replacing continuous derivatives by Jackson
symbols immediately leads to necessity to introduce the quantum plane and
$q$-differential calculus \cite{wz}
\label{e11}
\[ xy=qyx\ , \]
\[ D_x x=1+q^2 xD_x+(q^2-1) yD_y \quad ,\quad D_x y=qyD_x \ ,\]
\[ D_y x=qxD_y\quad , \quad D_y y=1+q^2 yD_y \ ,\]
\begin{equation}
D_xD_y=q^{-1}D_yD_x \ .
\end{equation}
The formulae analogous to (4) have the form
\label{e12}
\[ {(\tilde J^1_2 (n))}^{n+1} \equiv
(x^2 D_x\ +\ xy D_y - \{ n\} x)^{n+1} = \]
\begin{equation}
\sum_{k=0}^{k=n+1} q^{2n^2-n(k-2)+k(k-1)} {n+1 \choose k}_q x^{2n+2-k}y^k
D_x^{n+1-k} D_y^k  \ ,
\end{equation}
where
\[ {n \choose k}_q \equiv {\{n\}! \over {\{k\}!\{n-k\}!}}\ ,\ \{n\}! =
\{1\} \{2\}\ldots \{n\} \]
are $q$-binomial coefficient and $q$-factorial, respectively. Like all
previous cases, if $y \in {\bold C}$, the operator $\tilde J^1_2 (n)$ is
one of generators of
$q$-deformed algebra $sl_3(\bold C)_q$ of finite-difference
operators, acting on the quantum plane and having the linear space
${\cal P}_n(x,y)=Span\{x^iy^j: 0 \leq i+j \leq n\}$
as a finite-dimensional representation; the l.h.s. of (12) annihilates
${\cal P}_n(x,y)$. If $y$ is Grassmann variable,  $\tilde J^1_2 (n)$ is a
generator of the $q$-deformed superalgebra $osp(2,2)_q$ possessing
finite-dimensional representation (see e.g. \cite{t1}).

As it has been done before (see (2), (3), (6), (7), (10)), the identity (12)
can be rewritten in an abstract form replacing $x,y$ and $D_x,D_y$ by abstract
operators obeying relations (11).

Introducing a quantum hyperplane \cite{wz}, one can generalize
the whole family of the operator identities (5)-(6) replacing continuous
derivatives by finite-difference operators and then by abstract operators,
obeying a certain $q$-deformed Heisenberg algebra.

One of us (A.T.) wants to express a deep gratitude to
Profs. M. Gromov, L. Michel, R. Thom and IHES, Bures-sur-Yvette,
and to Prof. F. Pham and the University of Nice
for kind hospitality and their interest to the present work, and also
to Prof. R. Askey for valuable discussion of operator identities.

\end{document}